\documentclass[nofootinbib,twocolumn,preprintnumbers]{revtex4-1}
\usepackage{amsmath}
\usepackage{amssymb}
\usepackage{graphicx}
\usepackage{booktabs}
\usepackage{color}

\newcommand{\as}{\alpha_s}

\newcommand{\pt}{{p_t^{\tiny{\mbox{H}}}}}
\newcommand{\mh}{m_{\tiny{\mbox{H}}}}
\newcommand{\mur}{\mu_{\tiny{\mbox{R}}}}
\newcommand{\muf}{\mu_{\tiny{\mbox{F}}}}

\newcommand{\kto}{k_{t,1}}
\newcommand{\kti}{k_{t,i}}
\newcommand{\dki}{\langle dk_{i}\rangle}
\newcommand{\dko}{\langle dk_{1}\rangle}

\begin{document}

\title{A new approach to the Higgs transverse-momentum resummation at NNLL+NNLO}

\preprint{LAPTH-017/16, OUTP-16-07P}

\author{Pier Francesco Monni$^1$, Emanuele Re$^2$, Paolo Torrielli\,$^3$\vspace{1em}}

\affiliation{$^1$ Rudolf Peierls Centre for Theoretical
  Physics,University of Oxford, Keble Road, Oxford OX1 3NP, UK}
\affiliation{$^2$ LAPTh, Univ. de Savoie, CNRS, B.P.110, Annecy-le-Vieux F-74941, France} 
\affiliation{$^3$ Dipartimento di Fisica, Universit\`a di Torino and INFN, Sezione di Torino, Via P. Giuria 1, I-10125, Turin, Italy}

\begin{abstract}
We propose a new approach to the resummation of the transverse-momentum distribution of a high-mass colour-singlet system in hadronic collisions. The resummation is performed in momentum space and is free of kinematic singularities at small transverse momentum. We derive a formula accurate at the next-to-next-to-leading-logarithmic level, and present the first matched predictions to next-to-next-to-leading order for Higgs-boson production in gluon fusion at the LHC. This method can be adapted to all observables featuring kinematic cancellations in the infrared region.
\end{abstract}


\maketitle

The determination of the properties of the scalar resonance discovered
in 2012 by ATLAS and CMS \cite{Aad:2012tfa,Chatrchyan:2012xdj} is
central to the entire LHC physics programme. At Run II, owing to the
increased collision energy and luminosity, the Higgs-boson production
rate will increase significantly. As a consequence, not only will the
analyses already performed benefit from the increase in
statistics, but soon it will become possible to study kinematic
distributions in detail.
Obtaining as accurate predictions as possible
for the Higgs differential spectra is crucial, especially in view of the fact
that the signal significance is very commonly optimised by
categorising candidate events according to their kinematic properties;
therefore only by means of precise predictions for the Higgs
distributions can the increased statistics be fully exploited to
extract physics results.

Among the Higgs production channels, the gluon-fusion mode is the
dominant one at the LHC; among the most relevant kinematic
distributions, the Higgs transverse momentum $\pt$ will be
increasingly important in analysing the forthcoming experimental
results.

In gluon fusion the Higgs $\pt$ is defined as the inclusive vectorial
sum over the transverse momenta of the recoiling QCD partons radiated
off the incoming gluons.
The fixed-order perturbative description of its differential
distribution features large logarithms in the form
$\as^n \ln^m(\mh/\pt)/\pt$, with $m\leq 2n-1$, which spoil the
convergence of the series at small $\pt$.
In order to obtain meaningful predictions in that phase-space region,
such terms must be resummed to all orders in $\as$, so that the
perturbative series can be recast in terms of dominant all-order
towers of logarithms. The logarithmic accuracy is commonly defined at
the level of the {\it logarithm} of the cumulative cross section,
where one refers to the dominant terms $\alpha_s^n \ln^{n+1}(\mh/\pt)$
as leading logarithms (LL), to terms $\alpha_s^n \ln^{n}(\mh/\pt)$ as
next-to-leading logarithms (NLL), to $\alpha_s^n \ln^{n-1}(\mh/\pt)$
as next-to-next-to-leading logarithms (NNLL), and so on.

Such logarithms of the ratio $\mh/\pt$ have been resummed up to NNLL
order in~\cite{Bozzi:2003jy,Bozzi:2005wk} using the formalism
developed in~\cite{Collins:1984kg,Catani:2000vq}, and
in~\cite{Becher:2010tm} using an effective-theory approach. A careful
study of the theoretical uncertainties has been also carried out
in~\cite{Neill:2015roa}, and a formalism to perform a joint
$\pt$/small-{\it x} resummation has been presented
in~\cite{Marzani:2015oyb}.
The differential distribution in fixed-order perturbation theory has
been known for several years at next-to-leading order
(NLO)~\cite{deFlorian:1999zd,Ravindran:2002dc,Glosser:2002gm}, and has
been recently improved through the computation of the
Higgs-plus-one-jet cross section at next-to-next-to-leading order
(NNLO)~\cite{Boughezal:2015dra,Boughezal:2015aha,Caola:2015wna,Chen:2016vqn}.
The inclusive cross section was computed at NNLO in
refs.~\cite{Harlander:2002wh,Anastasiou:2002yz,Ravindran:2003um} and
recently at next-to-next-to-next-to-leading order (N$^3$LO)
in~\cite{Anastasiou:2015ema,Anastasiou:2016cez}.
These results can be matched to a NNLL resummation in order to
obtain a prediction which is accurate over the whole $\pt$ spectrum,
analogously to what has been done for the leading-jet transverse
momentum in ref.~\cite{Banfi:2015pju}.

All of the aforementioned resummations rely on an
impact-parameter-space formulation, which is motivated by the fact
that the observable naturally factorises in this space as a product of
the contributions from each individual emission.
Conversely, in $\pt$ space one is unable to find, at a given order
beyond LL, a closed analytic expression for the resummed distribution
which is simultaneously free of any logarithmically subleading
corrections and of singularities at finite $\pt$
values~\cite{Frixione:1998dw}.
This fact has a simple physical origin: the region of small $\pt$
receives contributions both from configurations in which each of the
transverse momenta of the radiated partons is equally small (Sudakov
limit), and from configurations where $\pt$ tends to zero owing to
cancellations among non-zero transverse momenta of the emissions. The
latter mechanism becomes the dominant one at small $\pt$ and, as a
result, the cumulative cross section in that region vanishes as
${\cal O}(\pt^2)$ rather than being exponentially
suppressed~\cite{Parisi:1979se}.
If these effects are neglected in a resummation performed in
transverse-momentum space, the latter would feature a geometric
singularity at some finite value of $\pt$.

In this letter we propose a new method that solves the problem in
transverse-momentum space and extends the framework of
refs.~\cite{Banfi:2004yd,Banfi:2014sua} to treat all observables
affected by the aforementioned kinematic cancellations. We obtain a
NNLL-accurate formula for $\pt$ and match the result to the NNLO
differential distribution for the first time.
\\

The starting point for a NLL resummation is to consider an ensemble
of partons $k_1,\cdots,k_n$, independently emitted off the initial-state
gluons $\ell=1,2$. Momenta $k_i$ are parametrised
as $k_i = z_i^{(1)}p_1 + z_i^{(2)}p_2 + \kappa_{t,i}$, where $p_{1,2}$
are the momenta of the incoming gluons and $\kappa_{t,i}$ is a
space-like four-vector, orthogonal to $p_1$ and $p_2$,
i.e. $\kappa_{t,i}=(0,\vec{k}_{t,i})$, such that
$\kappa_{t,i}^2 = - \kti^2$.  By singling out the largest-$k_t$
emission (labelled by $\kto$), the cumulative cross section can be
written as
\vspace{-2mm}
\begin{align}
  \label{eq:Sigma-start}
  &\Sigma(\pt)=
    \int_0^{\pt}\!\! dp_t' \frac{d \sigma(p_t')}{dp_t'}=\sigma_0\int_0^{\infty}\!\!
    \langle dk_1\rangle R'(\kto)
    e^{-R(\epsilon \kto)}\notag\\&\hspace{3mm}\times\sum_{n=0}^{\infty} \frac{1}{n!} 
                                   \prod_{i=2}^{n+1}\int_{\epsilon
                                   \kto}^{\kto}\langle dk_i\rangle
                                   R'(\kti)\Theta\Big(\pt-|\vec{q}_{n+1}|\Big),
\end{align}
where $\sigma_0$ denotes the Born cross section,
$\vec{q}_{n+1} = \sum_{j=1}^{n+1}\vec{k}_{t,j}$, and the measure
$\dki$ is defined in eq.~\eqref{eq:matrix-element} below. The radiator
$R(\epsilon k_{t})$ reads~\cite{Banfi:2012yh}
\begin{align*}
 R(\epsilon k_{t}) &= 4 \int_{\epsilon k_t}^{\mh}
    \frac{dk'_t}{k'_t}
    \left(\frac{\as^{\tiny{\mbox{CMW}}}(k'_{t})}{\pi} C_A\ln \frac{\mh}{k'_t}-\as(k'_{t})\beta_0\right),                     
\end{align*}
where $\as^{\tiny{\mbox{CMW}}}(k_{t}) $ in the double-logarithmic part
indicates that the strong coupling is expressed in the CMW
scheme~\cite{Catani:1990rr}, which ensures the correct treatment of
non-planar soft corrections at NLL accuracy in processes with two hard
emitters. The independent-emission amplitude squared and its phase
space are parametrised in eq.~\eqref{eq:Sigma-start}
as~\cite{Banfi:2004yd,Banfi:2014sua} \vspace{-2mm}
\begin{align}
  \label{eq:matrix-element}
[dk] M^2(k) =\frac{dk_{t}}{k_{t}} \frac{d\phi}{2\pi} R'(k_{t})\equiv\langle dk\rangle R'(k_t),
\end{align}
where $R'(k_{t})= -k_t d R(k_{t})/dk_{t}$. The bounds in the
$\langle dk_i\rangle$ integrals of eq.~\eqref{eq:Sigma-start} apply to
the $\kti$ variables, while all azimuthal integrals are evaluated in
the range $[-\pi,\pi]$.
In eq.~\eqref{eq:Sigma-start} the parton luminosity implicit in
$\sigma_0$ is evaluated at a fixed factorisation scale $\muf$, while a
complete NLL description requires a scale of the order of
$\kto$. Since this approximation is irrelevant for the present
discussion, we ignore it for the moment, and account for the proper
treatment of the luminosity only in the main result
(eq.~\eqref{eq:Sigma-NNLL}) of this letter.

The NLL transverse-momentum resummation then proceeds by expanding the
various $\kti$'s of eq.~\eqref{eq:Sigma-start} around the observable
$\pt$, retaining terms only up to NLL in the cumulative cross
section. This amounts to writing \vspace{-2mm}
\begin{align}
\label{eq:Rexpansion}
R(\epsilon \kto) &= R(\pt) + R'(\pt)\ln\frac{\pt}{\epsilon \kto}
                   +\cdots\,,\notag\\
R'(\kti) &= R'(\pt) + \cdots\,,
\end{align}
where neglected terms contribute at most to NNLL order in $\pt$
space. The second term in the expansion of $R(\epsilon \kto)$ plays
the role of virtual contribution, cancelling the infrared divergences
associated with the real emissions to all orders in $\as$. With these
replacements, eq.~\eqref{eq:Sigma-start} becomes \vspace{-1mm}
\begin{align}
  & \Sigma(\pt)
    =\sigma_0e^{-R(\pt)}\int_0^{\infty}\dko
    R'(\pt)\Big(\frac{\pt}{\kto}\Big)^{-R'(\pt)}
    \notag
  \label{eq:Sigma-tmp}
    \notag\\&\times\epsilon^{R'(\pt)}\sum_{n=0}^{\infty} \frac{1}{n!} 
              \prod_{i=2}^{n+1}\int_{\epsilon
              \kto}^{\kto}\dki
              R'(\pt)\Theta\Big(\pt-|\vec{q}_{n+1}|\Big),
\end{align}
which evaluates to
\vspace{-2mm}
\begin{align}
\label{eq:divergent-sigma}
 \Sigma(\pt)
    =\sigma_0 e^{-R(\pt)} e^{-\gamma_E R'(\pt)}\frac{\Gamma\left(1-R'(\pt)/2 \right)}{\Gamma\left(1+R'(\pt)/2 \right)}.
\end{align}
Eq.~\eqref{eq:divergent-sigma} reproduces the result of
ref.~\cite{Frixione:1998dw}; the geometric singularity at $R'(\pt)=2$
invalidates the result near the peak of the distribution, as a
consequence of the dominance of the aforementioned cancellation
mechanism over the usual Sudakov suppression. This comes about since
in the asymptotic limit $\pt\ll \kto$, the second line of
eq.~\eqref{eq:Sigma-tmp} scales as $(\pt/\kto)^2$ \cite{Banfi:2001bz},
which causes the cumulative cross section to diverge at $R'(\pt) = 2$.

The issue hides behind expansion~\eqref{eq:Rexpansion}, which was
performed with the aim of neglecting subleading effects: such an
expansion is indeed valid only in the region where $\pt/\kto\gtrsim1$,
while it cannot be applied when $\pt/\kto \to 0$.  A natural solution
can be achieved by using an impact-parameter-space
formulation~\cite{Dokshitzer:1978yd,Parisi:1979se}, since the
conjugate variable $b$ correctly
describes the vectorial nature of the $\pt\to 0$ limit.

However, the problem can also be overcome in direct space by simply
expanding $\kti$ around $\kto$ instead of $\pt$, namely \vspace{-2mm}
\begin{align}
R(\epsilon \kto) &= R(\kto) + R'(\kto)\ln\frac{1}{\epsilon} +
                     \cdots\,,\notag\\
\label{eq:R'expansion-fine}
R'(\kti) &= R'(\kto) + \cdots\,.
\end{align}
The resulting cumulative cross section reads
\vspace{-2mm}
\begin{align}
  & \Sigma(\pt)
    =\sigma_0\int_0^{\infty}\dko
    R'(\kto)
    e^{-R(\kto)}\epsilon^{R'(\kto)}\notag
  \label{eq:Sigma-fine}
\notag\\&\times
\sum_{n=0}^{\infty} \frac{1}{n!}
  \prod_{i=2}^{n+1}\int_{\epsilon
    \kto}^{\kto}\dki
  R'(\kto)\Theta\Big(\pt-|\vec{q}_{n+1}|\Big)\,.
\end{align}
Since by construction $\kti/\kto \leq 1$, the expansion in
\eqref{eq:R'expansion-fine} is always justified, and the exponential
factor regularises the $\pt/\kto \to 0$ limit.
Eq.~\eqref{eq:Sigma-fine} can be effectively interpreted as a
resummation of the large logarithms $\ln(\mh/\kto)$, and the
logarithmic order is defined in terms of the latter. This formulation
provides a correct description of both mechanisms that drive the limit
$\pt\to 0$, and it can be shown~\cite{WIP} that
eq.~\eqref{eq:Sigma-fine} reproduces the correct power-suppressed
scaling in this region~\cite{Parisi:1979se}. The corresponding formal
accuracy in terms of the logarithms $\ln(\mh/\pt)$ will be the same,
and the result differs from eq.~\eqref{eq:divergent-sigma} by
subleading logarithmic terms.

The above treatment can be systematically extended to NNLL. Since the
observable considered here is fully inclusive over QCD radiation,
the initial equation~\eqref{eq:Sigma-start} already contains most of
the NNLL contributions, as shown in ref.~\cite{Banfi:2014sua}.
More specifically, one should modify eq.~\eqref{eq:Sigma-fine}
introducing the NNLL radiator $R_{\rm NNLL}$ which is the same as for
the jet-veto resummation~\cite{Banfi:2012jm}, and retaining the next
term in the expansion~\eqref{eq:R'expansion-fine}, which involves the
second derivative of the radiator
$R''(\kto) \equiv -\kto d R'(\kto)/d\kto$; the parton luminosity is to
be evaluated at a scale of the order of $\kto$, as will be detailed
in~\cite{WIP}.
It is furthermore convenient to neglect N$^3$LL terms in the
$R'(\kto)$ and $R''(\kto)$ functions. We introduce the notation
\vspace{-2mm}
\begin{align}
R'(\kto) &= \hat{R}'(\kto) +\delta\hat{R}'(\kto) +\cdots\notag,\\
\label{eq:R'split}
R''(\kto) &= \hat{R}''(\kto) + \cdots,
\end{align}
where the functions $\hat{R}'(\kto)$ and $\delta\hat{R}'(\kto)$ are
NLL and NNLL, respectively, the neglected terms are at most of order
$\as^n\ln^{n-2} (\mh/\kto)$, and $\hat{R}''(\kto)$ indicates the
derivative of $\hat{R}'(\kto)$. The expressions for all these
functions can be found in the appendix of ref.~\cite{Banfi:2012jm}.
The NNLL cumulative cross section is thus conveniently recast as 
\begin{widetext}
\vspace{-4mm}
\begin{align}
  &\Sigma(\pt)=\int_0^{\infty}\dko\left[ \epsilon^{\hat{R}'(\kto)}\sum_{n=0}^{\infty} \frac{1}{n!}
  \prod_{i=2}^{n+1}\int_{\epsilon
    \kto}^{\kto}\dki
  \hat{R}'(\kto)\right]
    \bigg\{\partial_L\left[-e^{-R_{\rm NNLL}(\kto)}{\cal L}
    \right]
    \Theta\Big(\pt-|\vec{q}_{n+1}|\Big)
    \notag\\
  &+ e^{-R(\kto)}\hat{R}'(\kto)\!\int_{\epsilon
    \kto}^{\kto}\langle d k_{s}\rangle
  \left[  \left(
    \delta\hat{R}'(\kto)+\hat{R}''(\kto)
    \ln\frac{\kto}{k_{t,s}}\right) \hat{{\cal
    L}}- \partial_{L}\hat{{\cal
    L}}
    \right]
\label{eq:Sigma-NNLL}
      \!\bigg[\Theta\Big(\pt-|\vec{q}_{n+1,s}|\Big)
      - \Theta\Big(\pt-|\vec{q}_{n+1}|\Big)\bigg]\bigg\},
\end{align}
\end{widetext}
with $\vec{q}_{n+1,s}=\vec{q}_{n+1}+\vec{k}_{t,s}$. The above formula
can be evaluated by means of Monte Carlo methods.
The parton luminosity ${\cal L}$ is defined as
\vspace{-2mm}
\begin{align*}
{\cal L}&
  =\frac{\alpha^2_s(\mur)}{576 \pi v^2}\tau\sum_{i,j}\int_\tau^{1} \frac{dx_1}{x_1}
  \int_{x_1}^{1}\frac{dz_1}{z_1}\int_{\tau/x_1}^{1}\frac{dz_2}{z_2}\left[HC
  C\right]_{gg;ij}\notag\\
&\times f_i\!\left(x_1/z_1, e^{-L} \muf\right)f_j\!\left(\tau/x_1/z_2, e^{-L}
  \muf\right),
\end{align*}
where $\mur$ is the renormalisation scale, $\tau=\mh^2/s$, $v$ is the vacuum expectation value of the Higgs field, and
$L=\ln(Q/\kto)$, the resummation scale $Q$ being introduced as shown
in refs.~\cite{Bozzi:2005wk,Banfi:2012jm}.
The factor $\left[HCC\right]$ is defined as
\begin{align}
  &\left[HCC\right]_{gg;ij} =
  H^{H}_{g}(\alpha_s(\mur),\mur,Q,\mh)\notag\\
&\times\big[ C_{g i}(z_1;
  \alpha_L, \mur, \muf, Q) C_{g j}(z_2;
  \alpha_L, \mur, \muf,
  Q)\notag\\
\label{eq:HCC}
&+G_{g i}(z_1;
  \alpha_L, \mur, \muf) G_{g j}(z_2;
  \alpha_L, \mur, \muf)\big],
\end{align}
where $\alpha_L = \alpha_s(\mur)/(1-2 \alpha_s(\mur)\beta_0 L)$. The
product in eq.~\eqref{eq:HCC} is further expanded neglecting constant
terms beyond ${\cal O}(\as^2)$. The functions $H^H_g$, $C_{ij}$ and
$G_{ij}$ are deduced using the results of ref.~\cite{Catani:2011kr},
after including the proper scale dependences.  The NLL luminosity
$\hat{\cal L}$ is obtained from $\cal L$ by setting
$\left[HCC\right]_{gg;ij}=\delta_{gi}\delta_{gj}\delta(1-z_1)\delta(1-z_2)$,
and it reproduces the Born cross section $\sigma_0$ in the limit
$L\to 0$.

The various contributions to eq.~\eqref{eq:Sigma-NNLL} are described
in what follows.  The first line includes all NNLL corrections to the
hardest emission $k_1$; this reflects the arguments that led to
eq.~\eqref{eq:Sigma-fine}, properly extended to NNLL. The corrections
to the remaining emissions are encoded in the second line of
eq.~\eqref{eq:Sigma-NNLL}, where only a single emission $k_s$ of the
ensemble is corrected. This is implemented in
eq.~\eqref{eq:Sigma-NNLL} by computing the difference between the
observable evaluated using all emissions, including the modified one
$k_s$, and the observable obtained by neglecting $k_s$.
Since these configurations give at most a NNLL correction, it suffices
to use the NLL luminosity $\hat{{\cal L}}$ and radiator $R(\kto)$ in
the second line of eq.~\eqref{eq:Sigma-NNLL}.
%
Finally, the term proportional to $\partial_L\hat{{\cal L}}$ accounts
for the luminosity contribution to the NNLL hard-collinear correction,
described by a DGLAP-evolution step between $\epsilon\kto$ and $\kto$.
The corresponding contribution for the first emission is encoded in the
first line of eq.~\eqref{eq:Sigma-NNLL}, where terms beyond NNLL are
included in order to reproduce the exact differential for the
hardest-emission probability. The latter are physical contributions,
namely they are a subset of the subleading terms that would be
generated by retaining higher orders in the resummation.

As a check of eq.~\eqref{eq:Sigma-NNLL}, we have expanded it around
$\kto=\pt$, neglecting N$^3$LL terms in $\ln(\mh/\pt)$. This
approximation is the source of the singularity in
eq.~\eqref{eq:divergent-sigma}, but it contains all the correct NNLL
terms at a given fixed order in $\as$, which can be used as a powerful
test of the accuracy of our result. Formula~\eqref{eq:Sigma-NNLL}
reproduces the analytic result
reported in the appendix of ref.~\cite{Banfi:2012jm} at NNLL.\\
%

As a phenomenological application of eq.~\eqref{eq:Sigma-NNLL} we
perform a matching to the N$^3$LO cumulant, which is obtained by
combining the total N$^3$LO cross section~\cite{Anastasiou:2016cez}
and the NNLO Higgs-plus-jet cross section~\cite{Boughezal:2015dra}. We
perform an additive matching, and unitarity at high $\pt$ is restored by introducing the modified logarithms
\begin{equation*}
\ln(Q/\kto)\to 1/p\ln\big[(Q/\kto)^p+1\big],
\end{equation*}
where we choose $p=2$.\footnote{This choice is made only for
  consistency with the literature we compare with. A study on the
  optimal choice of $p$ is left for future work.}  We consider
$13$\,TeV LHC collisions, with $\mh=125$\,GeV, and {\tt PDF4LHC15}
\cite{Butterworth:2015oua} parton densities at NNLO. The central
prediction uses $\mur=\muf=\mh$, and $Q=\mh/2$. The perturbative
uncertainty for all predictions is estimated by varying both $\mur$
and $\muf$ by a factor of two in either direction while keeping
$1/2\leq \mur/\muf \leq 2$.  Moreover, for central $\mur$ and $\muf$
scales we vary the resummation scale $Q$ by a factor of two in either
direction.
%

\begin{figure}[t!]
  \centering
  \includegraphics[width=0.9\columnwidth]{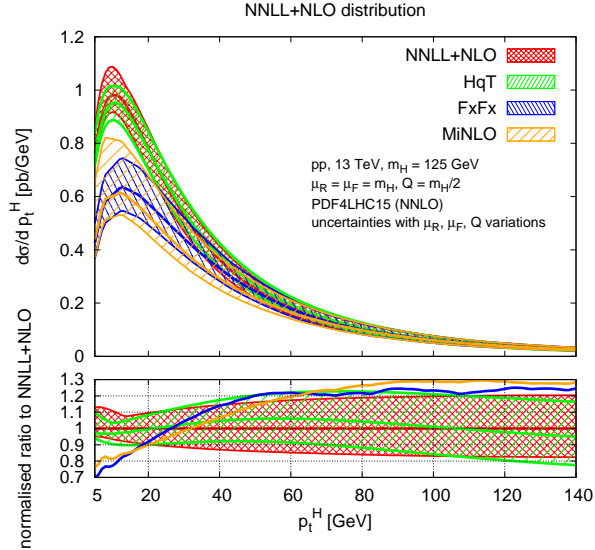} 
  \caption{\footnotesize{Comparison of the Higgs $\pt$ NNLL+NLO
      prediction as obtained in this letter (red) to \texttt{HqT}
      (green). For reference, the predictions obtained with MiNLO at
      NLO (orange), and FxFx (blue) are shown. Lower panel: ratio of
      the various distributions, normalised to their respective
      central-scale inclusive cross sections, to the central NNLL+NLO
      prediction. Uncertainty bands are shown only for the
      resummed results.}}
  \label{fig:results1}
  \end{figure}
\begin{figure}[t!]
  \centering
  \includegraphics[width=0.9\columnwidth]{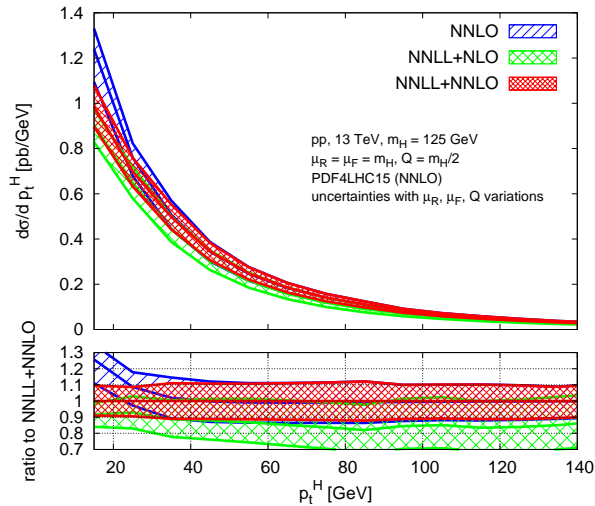}  
  \caption{\footnotesize{Higgs $\pt$ at NNLL+NNLO (red), NNLL+NLO
    (green), and NNLO (blue). Lower panel: ratio of the
    three predictions to the NNLL+NNLO one.}}
  \label{fig:results2}
  \end{figure}
  \noindent To validate our result, in the main panel of figure~\ref{fig:results1} we
  show the comparison of our prediction for the
  Higgs-transverse-momentum spectrum at NNLL+NLO to that obtained with
  \texttt{HqT}~\cite{Bozzi:2005wk,deFlorian:2011xf}.
  As expected, we observe a very good agreement over the entire $\pt$
  range between these two results, which have the same perturbative
  accuracy. Our NNLL+NLO prediction is moderately higher in
  the peak of the distribution, and lower at intermediate
  $\pt$ values, although this pattern may slightly change with
  different central-scale choices. These small differences have to do
  with the different treatment of subleading effects in the two
  resummation methods.  
 The agreement of the two results, both for the central scale and for the
  uncertainty band, is even more evident in the lower inset of figure~\ref{fig:results1},
  which displays the ratio of the various distributions, each normalised to its central-scale 
  inclusive rate, to our normalised central NNLL+NLO curve.

  For comparison, figure~\ref{fig:results1} also reports the $\pt$
  distribution obtained with the NLO version of \texttt{POWHEG}+MiNLO
  \cite{Nason:2004rx,Alioli:2010xd,Hamilton:2012rf}, and with the
  \texttt{MadGraph5\_aMC@NLO}+FxFx
  \cite{Frixione:2002ik,Alwall:2014hca,Frederix:2012ps} event
  generators, using default renormalisation and factorisation
  scales for the two methods (in FxFx a merging scale $\mu_Q=\mh/2$
  has been employed). Both generators are interfaced to
  \texttt{Pythia\,8.2} \cite{Sjostrand:2014zea}, without including
  hadronisation, underlying event, and primordial $k_{\perp}$ (whose
  impact has been checked to be fully negligible for this observable),
  and use {\tt PDF4LHC15} parton densities at NLO. By inspecting the
  normalised ratios shown in the lower panel, one observes that the
  shape of the Monte-Carlo predictions deviates significantly
  from the NNLL+NLO results at $\pt \lesssim 60$\,GeV.

  Figure~\ref{fig:results2} shows the comparison of the matched
  NNLL+NNLO result to the NNLL+NLO and the fixed-order NNLO
  predictions.
  The inclusion of the NNLO corrections leads to a $10-15\%$ increase in
  the matched spectrum for $\pt > 15$\,GeV, and to a consistent reduction
  in the perturbative uncertainty, to the $\pm10\%$-level in the considered $\pt$
  range.
  The impact of resummation on the fixed order becomes
  increasingly important for $\pt \lesssim 40$\,GeV, reaching about $25\%$ at
  $\pt = 15$\,GeV.  For $\pt \gtrsim 40$\,GeV, the
  matched prediction reduces to the NNLO one.\\

  In this letter we have presented a new method, entirely formulated
  in momentum space, for the resummation of the transverse momentum of
  a colour-singlet final state in hadronic collisions. We have used it
  to obtain the first NNLL+NNLO prediction for the Higgs-boson
  transverse-momentum spectrum at the LHC. Higher-order logarithmic
  corrections beyond NNLL can be systematically included within this
  framework.
  Our approach does not rely on any specific factorisation theorem,
  and therefore it can be generalised to treat any observable
  featuring kinematic cancellations in the infrared region -- like for
  instance $\phi^*$ in Drell-Yan pair production \cite{Banfi:2010cf}
  or the oblateness in electron-positron annihilation -- as well as to
  compute any other observable which can be treated with the methods
  of refs.~\cite{Banfi:2004yd,Banfi:2014sua}.
  Notably, this paves the way for formulating a simultaneous
  resummation for the Higgs and the leading-jet transverse momenta at
  NNLL.\\

  We are very grateful to F.~Caola for providing us with the NNLO
  distributions used in this work and for checking the fixed-order
  results, and to G.~Salam for very useful discussions on some aspects
  of the topic treated here, as well as for help with the use of the
  \texttt{HOPPET} code~\cite{Salam:2008qg}. We also thank F.~Dulat for
  providing us with the N$^3$LO total cross sections used in the
  results, and A.~Banfi, L.~Magnea and G.~Zanderighi for carefully
  reading the manuscript and for providing valuable comments. We
  acknowledge the hospitality of CERN's Theory Department while part
  of this work was carried out.
  PM was partly supported by the Swiss National Science Foundation
  (SNF) under grant PBZHP2-147297. PM and ER have benefitted from the
  ERC grant 614577 HICCUP. The work of PT has received funding
  from the European Union Seventh Framework programme for research and
  innovation under the Marie Curie grant agreement N. 609402-2020
  researchers: Train to Move (T2M).

\end{document}